\documentclass[sigconf]{acmart}
\AtBeginDocument{%
  }

\setcopyright{acmlicensed}

\copyrightyear{2025}
\acmYear{2025}
\setcopyright{rightsretained}
\acmConference[APNET 2025]{9th Asia-Pacific Workshop on Networking}{August 07--08, 2025}{Shang Hai, China}
\acmBooktitle{9th Asia-Pacific Workshop on Networking (APNET 2025), August 07--08, 2025, Shang Hai, China}
\acmPrice{}
\acmDOI{10.1145/3735358.3735383}
\acmISBN{979-8-4007-1401-6/25/08}

\acmSubmissionID{262}

\usepackage{xcolor}  

\usepackage{graphicx}
\usepackage{subcaption}
\usepackage{float}

\usepackage{pifont}
\newcommand{\cmark}{\ding{51}}%
\newcommand{\xmark}{\ding{55}}%

\begin{document}

\title{PromptMobile: Efficient Promptus for Low Bandwidth Mobile Video Streaming}

\author{
       {
       Liming Liu,
       Jiangkai Wu,
       Haoyang Wang, 
       Peiheng Wang,
       Zongming Guo,
       Xinggong Zhang
       }\\
       \textit{Peking University}\\
       {llm@stu.pku.edu.cn}
}


\begin{abstract}
 Traditional video compression algorithms exhibit significant quality degradation at extremely low bitrates. Promptus emerges as a new paradigm for video streaming, substantially cutting down the bandwidth essential for video streaming. However, Promptus is computationally intensive and can not run in real-time on mobile devices. This paper presents PromptMobile, an efficient acceleration framework tailored for on-device Promptus. Specifically, we propose (1) a two-stage efficient generation framework to reduce computational cost by 8.1x, (2) a fine-grained inter-frame caching strategy to reduce redundant computations by 16.6\%, (3) system-level optimizations to further enhance efficiency. The evaluations demonstrate that compared with the original Promptus, PromptMobile achieves a 13.6x increase in image generation speed. Compared with other streaming methods, PromptMobile achieves an average LPIPS improvement of 0.016 (compared with H.265), reducing 60\% of severely distorted frames (compared to VQGAN). 
\end{abstract}




\maketitle
\pagestyle{plain}

\section{INTRODUCTION}

Currently, video traffic accounts for over 65\% of the total Internet traffic~\cite{global-phenomena24}, making it a dominant component of network transmission. Traditional video compression algorithms like H.264\cite{264}, H.265\cite{265}, and VP9\cite{mukherjee2015technical} exhibit significant quality degradation at extremely low bitrates. Recently, Promptus\cite{wu2024promptus} emerges as a new paradigm for video streaming. At the sender, it leverages Stable Diffusion\cite{sd} (a large text2image model) to inversely generate semantic prompts from videos. At the receiver, Stable Diffusion re-generates video frames based on the received prompts. The reconstructed videos preserve per-pixel fidelity to the original videos. The data size of prompts is far smaller compared to that of videos. So Promptus substantially cuts down the bandwidth essential for video transmission.

However, Stable Diffusion is computationally intensive. For instance, it requires high-performance desktop-grade GPUs (such as the NVIDIA 4090) to achieve real-time image generation. Currently, more than 70\% of American Internet users rely on mobile devices for video streaming\cite{statista_mobile_video}, yet their limited computational resources prevent Stable Diffusion from running in real-time, posing a significant challenge to the deployment of Promptus.

\begin{figure}[t]  
    \centering        
    \setlength{\abovecaptionskip}{2.mm}
    \includegraphics[width=0.44\textwidth]{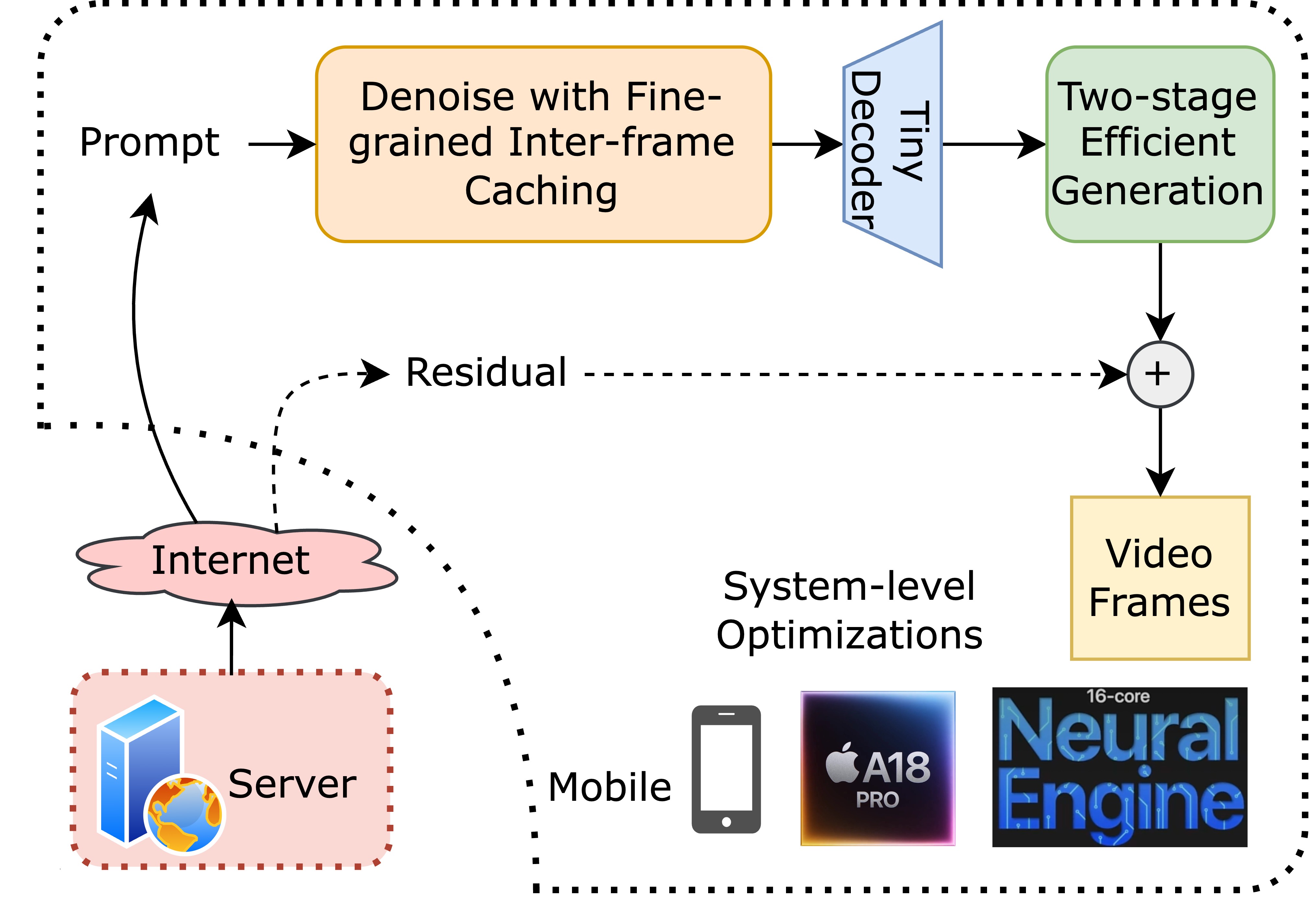}
    \caption{System overview of PromptMobile. Prompts are first transmitted from the server to the mobile client. The client uses a UNet network, optimized for the Apple Neural Engine (ANE), to perform single-step denoising and generate a low-resolution latent representation. This UNet also integrates an inter-frame caching strategy to reduce redundant computation. A TinyDecoder then reconstructs a stitched low-resolution image from the latent representation, which is subsequently unstitched and upsampled by a two-stage generation module to produce high-resolution frames. An optional residual stream can be applied to further enhance visual quality when needed. }  
    \label{fig:intro-pipeline}  
    \vspace{-5mm}
\end{figure}

Extensive research efforts have been dedicated to accelerating diffusion models on mobile platforms. 

First, Google\cite{chen2023speed} presents a series of implementation optimizations to enhance inference speed, achieving up to 5 FPS on an iPhone 14 Pro Max. However, this remains insufficient for real-time video generation.

Second, the edge-cloud collaborative approach\cite{xu2024accelerating} partitions the denoising and decoding processes between the mobile device and cloud server, which reduces the amount of computation that the mobile device needs to perform. However, this method requires transmitting massive intermediate variables, which introduces unacceptable bandwidth overhead.

Third, some methods\cite{wimbauer2024cache} design a cache strategy between the denoising steps. They predict the result similarity with the input similarity, directly reusing some results of previous steps to avoid redundant computations. However, these prediction-based caching techniques usually have low accuracy, leading to a in image quality. Additionally, these caching mechanisms are designed for multi-step denoising, but Promptus adopts a single-step denoising approach.

\textbf{In this paper, our vision is to run Promptus over 30 frames per second (FPS) on mobile devices, enabling low-bitrate on-device video streaming.}
To achieve this, we propose \textbf{PromptMobile}, an efficient acceleration framework tailored for on-device Promptus. Specifically:


\textbf{Two-stage efficient generation}. We propose a hierarchical generation framework that reduces computational cost by 8.1$\times$. It first uses the diffusion model to generate low-resolution images, which are then upscaled by a lightweight upsampling module. To improve the quality of low-resolution outputs without increasing overhead, we introduce an image-stitching technique (\S \ref{sec:method-twostage}).


\textbf{Fine-grained inter-frame caching.} We introduce an inter-frame caching strategy that takes full advantage of the inter-frame similarity to reduce redundant computations. First, to maintain image quality when caching, we proposed a lossless cache (\S \ref{sec:method-cache}). Second, to further reduce inference latency, we develop a lossy cache (\S \ref{sec:method-cache}). With those two caching strategies, we effectively decrease the computational complexity by 16.6\%, while ensuring high fidelity.




\textbf{System-level optimizations.} We further enhance efficiency through various system-level improvements. First, to mitigate the quality degradation caused by the upsampling and caching, we incorporate a collaborative optimization method that integrates the two operations into the end-to-end prompt inversion training process. Second, if the video quality still remains unsatisfactory, we propose a residual enhancement approach, transmitting an extra residual stream to compensate for the quality degradation. Third, to reduce inference complexity, we replace the original decoder with a low-complexity Tiny Decoder. Finally, to further enhance inference efficiency, we develop an efficient implementation utilizing Apple Silicon's Neural Engine.

Our empirical evaluations demonstrate that PromptMobile achieves a 13.6$\times$ increase in image generation speed while maintaining the same visual quality and bandwidth overhead. Under a bandwidth constraint of 280kbps, our approach outperforms traditional video compression algorithms, achieving an average LPIPS improvement of 0.016 compared with H.265\cite{265}, reducing 60\% of severely distorted frames compared to VQGAN\cite{li2023reparo,esser2021taming}. 

The remainder of this paper is structured as follows: \S \ref{sec:motivation} introduces background on Promptus and mobile stable diffusion model acceleration, and points out our main insights that have potential gains. \S \ref{sec:mothod} presents the overview and details the implementation of our methodologies. \S \ref{sec:evaluation} presents the experimental evaluation. \S \ref{sec:limitations} discusses the limitations of our current approach, and \S \ref{sec:conclusion} concludes the paper. 

\section{MOTIVATION AND RELATED WORK}
\label{sec:motivation}

In this section, we first set up the background of Promptus (\S\ref{sec:promptus}), followed by a discussion of existing model acceleration techniques and their limitations (\S\ref{sec:related}). Then, we present the main insights behind our work (\S\ref{sec:moti}).

\vspace{-2mm}

\subsection{Promptus}
\label{sec:promptus}
Promptus\cite{wu2024promptus} is a novel video transmission pipeline that exploits the prior knowledge embedded in Stable Diffusion to achieve pixel-level high-fidelity video reconstruction under stringent bandwidth constraints. At the transmitter, Promptus first randomly initializes a prompt and then uses the Stable Diffusion to generate video frames. Next, it optimizes the prompt via gradient descent until a prompt for generating pixel-level aligned frames is obtained. Since only the prompt needs to be transmitted and the original frames can be restored using the Stable Diffusion at the client, this method significantly reduces the bandwidth for video transmission.

It is worth mentioning that Promptus employs the turbo version of Stable Diffusion, which is derived from fine-tuning the basic Stable Diffusion using an adversarial loss. This model enables a \textbf{single-step denoising} process to restore a reasonably high-quality image, significantly accelerating image generation. Additionally, Promptus achieves \textbf{intra-frame compression} by decomposing the $77 \times 1024$ prompt matrix into two lower-rank matrices: $U$ of size $77 \times \text{rank}$ and $V$ of size $\text{rank} \times 1024$. To further enhance efficiency, Promptus introduces an \textbf{inter-frame compression} strategy, which only trains a prompt for each keyframe, while the prompt for intermediate frames is obtained through linear interpolation between that of the two adjacent keyframes. 

However, real-world evaluations reveal that when running the Stable Diffusion 2.1 Turbo model with a 512 $\times$ 512 resolution on an iPhone 16 Pro Max, it can only reach 3.1 FPS. This indicates that it is challenging for Promptus to achieve real-time video generation on devices with limited computing power like mobile phones.

\subsection{Mobile Model Acceleration}
\label{sec:related}
Existing approaches for accelerating diffusion models on mobile devices primarily fall into three categories: 

\noindent\textbf{Engineering optimization}: Google\cite{chen2023speed} presents a series of implementation optimizations like FlashAttention to enhance inference speed, achieving up to 5 FPS on an iPhone 14 Pro Max. However, this remains insufficient for real-time video generation. Furthermore, modifications to the model architecture mandate a complete retraining process, which is too expensive in terms of computational resources and time to be acceptable. 

\noindent\textbf{Edge-Cloud collaboration}: MEG\cite{xu2024accelerating} partitions the denoising and decoding processes between the mobile device and cloud server, which reduces not only the amount of computation that the mobile device needs to perform but also the overall computation latency. However, this method depends on the cloud infrastructure and requires transmitting massive intermediate variables, which conflicts with Promptus's goal to minimize bandwidth usage.

\noindent\textbf{Cache-based methods}: This kind of method\cite{wimbauer2024cache} designs a cache strategy between multi-step denoising steps. They first predict the output similarity by input similarity. If the predicted similarity is large enough, they directly reuse the precomputed results to avoid redundant computations, thus reducing the inference cost significantly. However, these prediction-based caching techniques usually have low accuracy. With a large cache ratio, the image quality degrades substantially. Additionally, existing caching mechanisms, designed for multi-step denoising Stable Diffusion models, are not compatible with our single-step denoising model.

\subsection{Potential Gains and Insights}
\label{sec:moti}

\begin{figure}[t]
    \begin{minipage}[b]{0.48\linewidth}
        \centering
        \setlength{\abovecaptionskip}{0.mm}
        \includegraphics[width=\textwidth]{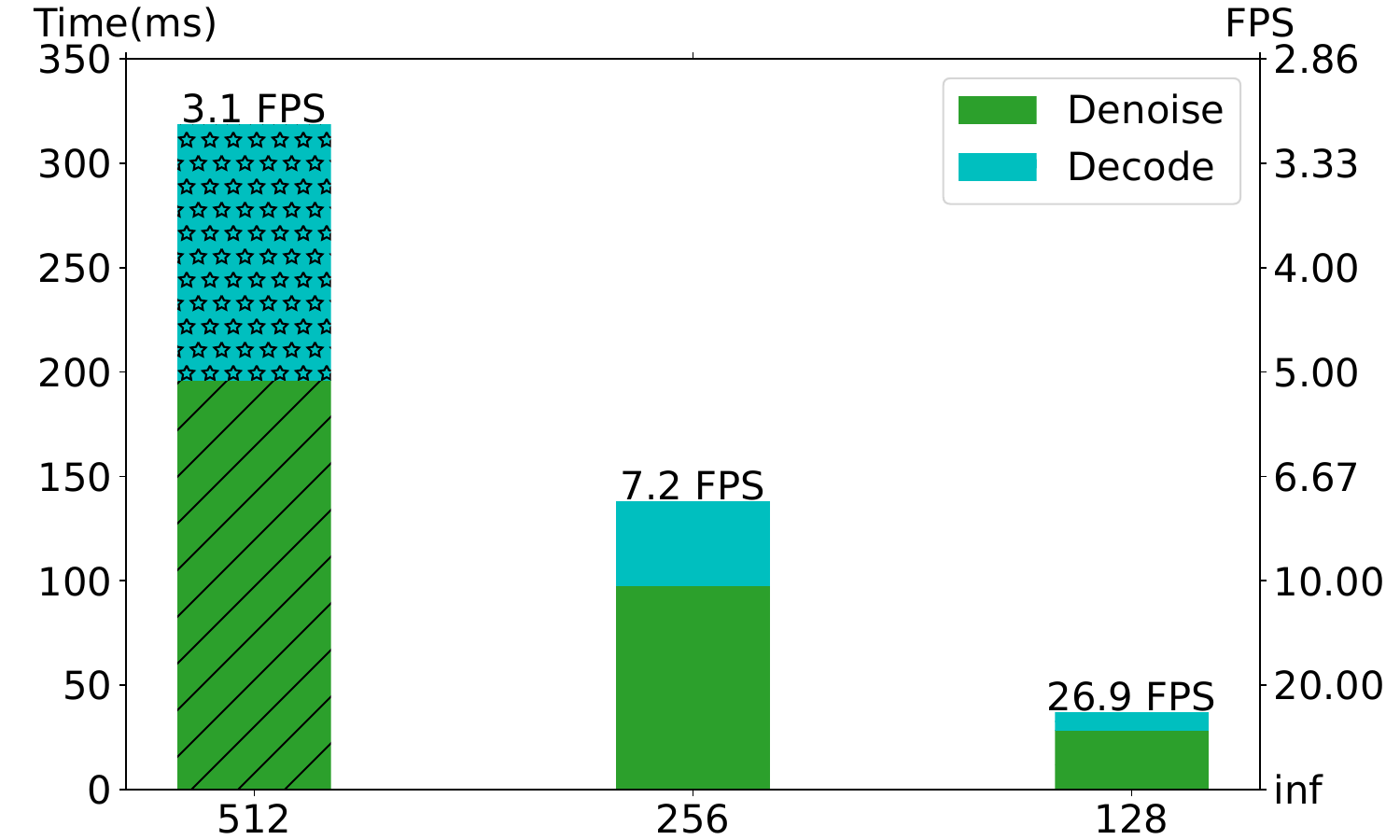}
        \caption{Lower resolutions lead to faster inference; 128×128 runs about 9 times faster than 512×512.}
        \label{fig:motivation-resolution-time}
    \end{minipage}
    \hfill
    \begin{minipage}[b]{0.48\linewidth}
        \centering
        \setlength{\abovecaptionskip}{0.mm}
        \includegraphics[width=\textwidth]{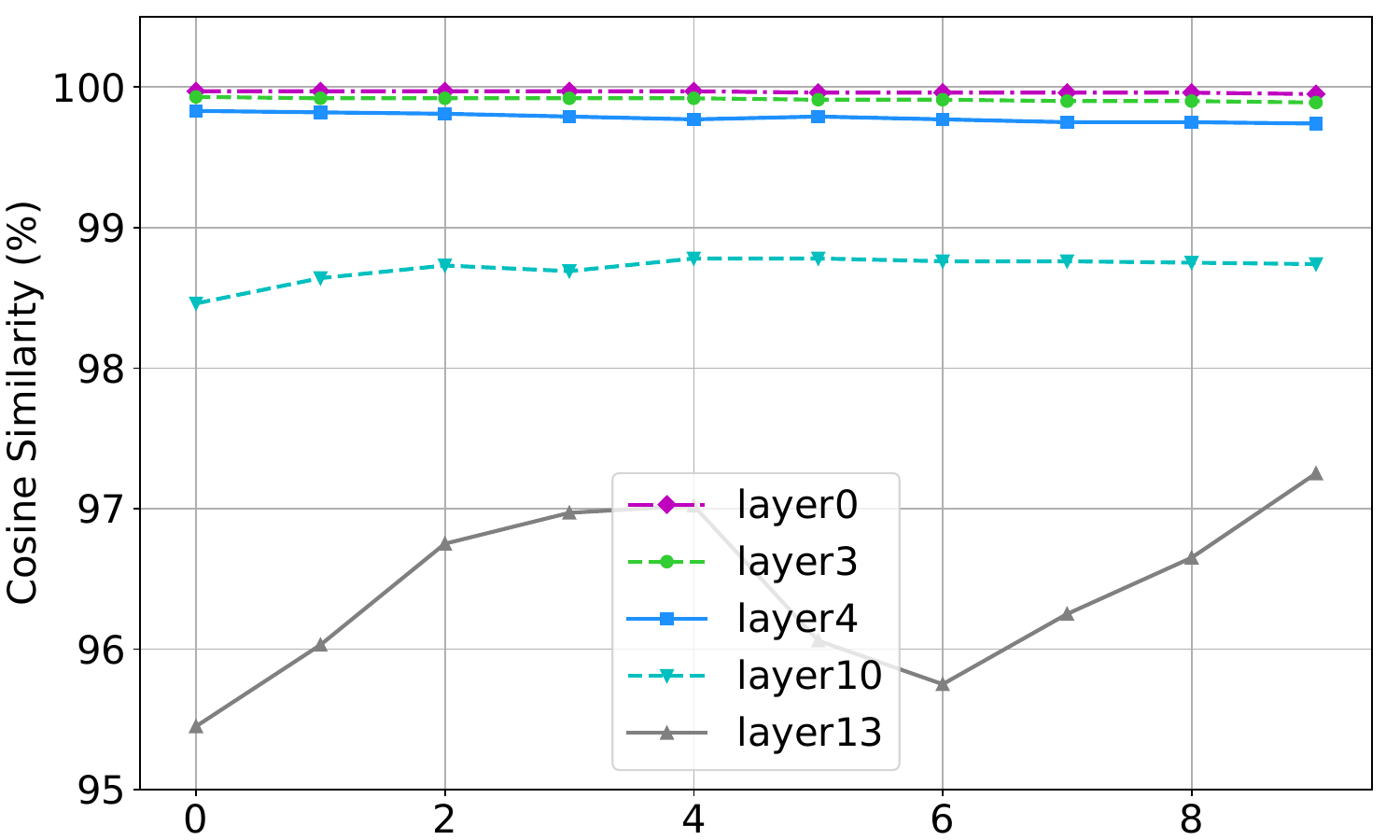}
        \caption{Prompt interpolation results in highly similar intermediate features across frames.}
        \label{fig:motivation-similarity}
    \end{minipage}
    \vspace{-4mm}
\end{figure}

Based on the above analysis, it is challenging to run Stable Diffusion for Promptus on mobile devices like an iPhone in real time. Fortunately, we have the following insights that have potential gains:

\noindent\textbf{Two-stage generation significantly reduces computational cost.} To accelerate diffusion model inference on mobile devices, we note that diffusion models can freely change the output resolution. As shown in Fig. \ref{fig:motivation-resolution-time}, the inference time is roughly proportional to the output image's pixel count. For instance, generating a 128 $\times$ 128 resolution image is about $9 \times$ faster than a 512 $\times$ 512 one. Considering the redundant information in images, we propose generating low-resolution images with the diffusion model and then using a lightweight upsampling module to obtain high-resolution, high-quality images. (\S \ref{sec:method-twostage})

\begin{figure}[t] 
\setlength{\abovecaptionskip}{0.mm}
    \centering
    \begin{subfigure}[b]{0.11\textwidth}
        \includegraphics[width=\textwidth]{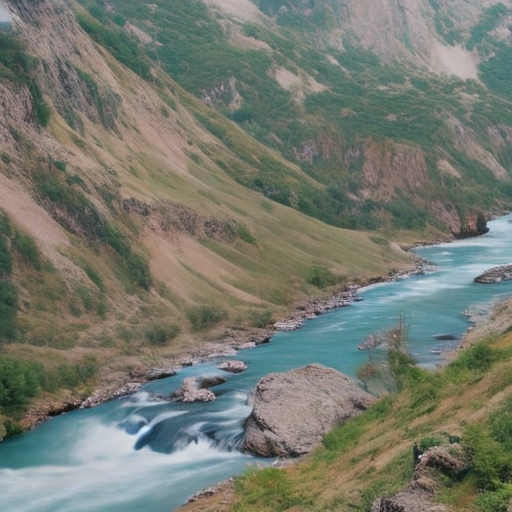}
        \caption{512 $\times$ 512}
        \label{fig:motivation-lowres-fig-512}
    \end{subfigure}
    \hfill
    \begin{subfigure}[b]{0.11\textwidth}
        \includegraphics[width=\textwidth]{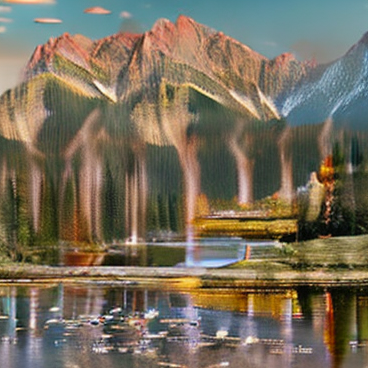}
        \caption{368 $\times$ 368}
        \label{fig:motivation-lowres-fig-368}
    \end{subfigure}
    \hfill
    \begin{subfigure}[b]{0.11\textwidth}
        \includegraphics[width=\textwidth]{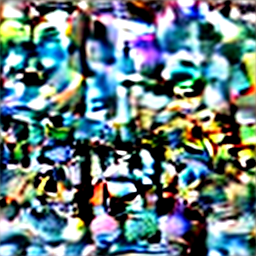}
        \caption{256 $\times$ 256}
        \label{fig:motivation-lowres-fig-256}
    \end{subfigure}
    \hfill
    \begin{subfigure}[b]{0.11\textwidth}
        \includegraphics[width=\textwidth]{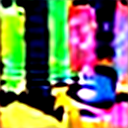}
        \caption{128 $\times$ 128}
        \label{fig:motivation-lowres-fig-128}
    \end{subfigure}
    \caption{The generated images becomes abstract when the resolution becomes lower.}
    \label{fig:motivation-lowres-fig}
    \vspace{-4mm}
\end{figure}

\begin{figure}[t] 
\setlength{\abovecaptionskip}{0.mm}
    \centering
    \begin{subfigure}[b]{0.11\textwidth}
        \includegraphics[width=\textwidth]{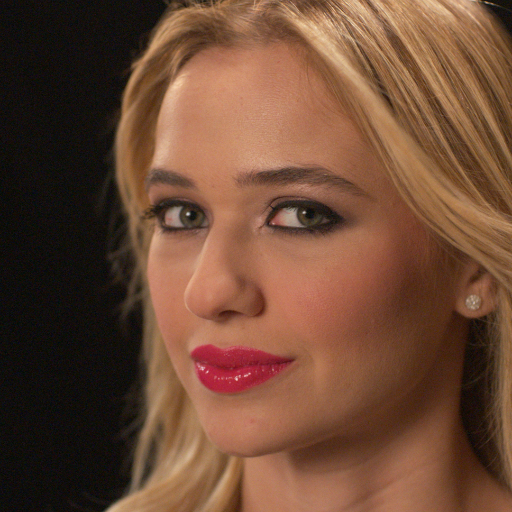}
        \caption{GT}
        \label{fig:motivation-lowuvg-fig-512}
    \end{subfigure}
    \hfill
    \begin{subfigure}[b]{0.11\textwidth}
        \includegraphics[width=\textwidth]{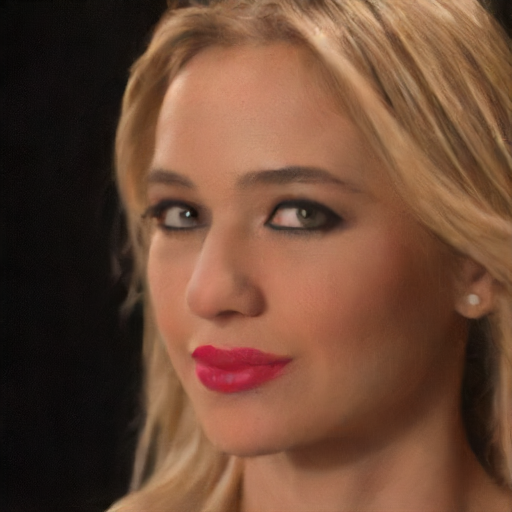}
        \caption{512 $\times$ 512}
        \label{fig:motivation-lowuvg-fig-368}
    \end{subfigure}
    \hfill
    \begin{subfigure}[b]{0.11\textwidth}
        \includegraphics[width=\textwidth]{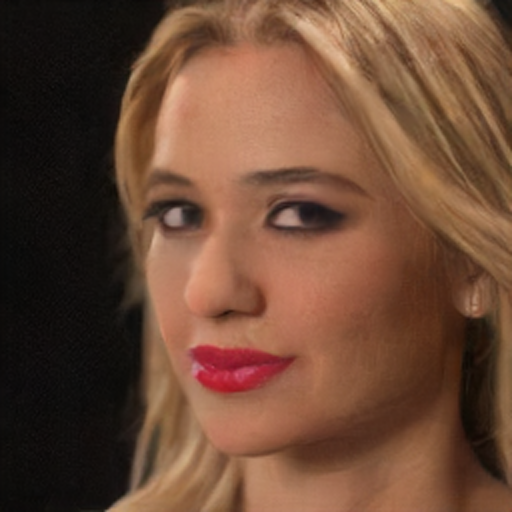}
        \caption{256 $\times$ 256}
        \label{fig:motivation-lowuvg-fig-256}
    \end{subfigure}
    \hfill
    \begin{subfigure}[b]{0.11\textwidth}
        \includegraphics[width=\textwidth]{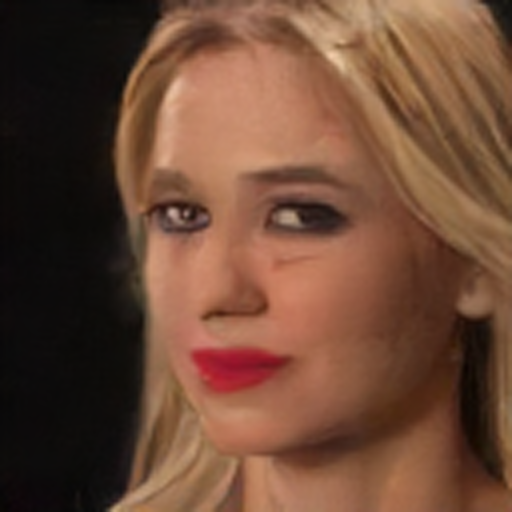}
        \caption{128 $\times$ 128}
        \label{fig:motivation-lowuvg-fig-128}
    \end{subfigure}
    \caption{The quality of the fitted image decreases as the resolution decreases.}
    \label{fig:motivation-lowuvg-fig}
    \vspace{-4mm}
\end{figure}

\noindent\textbf{Image stitching enables Stable Diffusion to generate low-resolution images.} Diffusion models lack pre-training on low-resolution images. As depicted in Fig. \ref{fig:motivation-lowres-fig}, when using the same prompt to generate images at decreasing resolutions, the output images become increasingly simplistic and abstract. Similarly, Fig. \ref{fig:motivation-lowuvg-fig} demonstrates that at lower resolutions, the images fitted by Promptus exhibit a significant decline in quality, with notable degradation in visual fidelity and a marked loss of fine details. To effectively fit low-resolution images without fine-tuning the diffusion model\cite{huggingface_minisd}, we propose an image-stitching approach where multiple adjacent low-resolution frames are combined to form a higher-resolution image, preserving fitting details while ensuring computational efficiency. (\S \ref{sec:method-twostage})

\noindent\textbf{Inter-frame caching to reduce computational overhead.} We note that in Promptus, intermediate prompts are derived by interpolating prompts from the adjacent keyframes. So some linear layers in the denoise network perform redundant computations across frames. So we can reduce the number of operations without any quality loss by caching key variables and using simple linear transformations. Additionally, Inspired by multi-step caching, we analyzed the similarity of intermediate variables in U-Net modules across adjacent frames using the human face videos from the UVG dataset. As Fig. \ref{fig:motivation-similarity} shows, intermediate representations between adjacent frames are relatively similar, like those across denoising timestamps, enabling inter-frame caching(\S \ref{sec:method-cache}). 

\section{METHODS}
\label{sec:mothod}

\subsection{Overview}


In response to the above insights, we design three modules for PromptMobile: Two-stage efficient generation, Fine-grained inter-frame caching, and System-level optimizations. The architecture of PromptMobile is shown in Fig. \ref{fig:intro-pipeline} and its workflow is as follows:

During video streaming, the server first prepares and transmits the prompts to the mobile client. Once the client receives the prompts, it first utilizes a denoise network that has been optimized for the \textbf{Apple Neural Engine (ANE)} and incorporates \textbf{Fine-grained inter-frame caching}. This denoise network performs a single-step denoising operation on the initial noise and generates a latent vector representing the image. Subsequently, a \textbf{TinyDecoder}, which has negligible processing time, is responsible for decoding the latent vector and output a low-resolution frame. Next, the \textbf{two-stage efficient generation} module is employed, upsampling the low-resolution frame to the final resolution. In addition, when necessary, the server will send an additional \textbf{Residual video stream} to the client, which contains supplementary information that can improve the quality of the generated frames. By combining the residual stream with the existing frames, the client can further enhance the quality of the restored video. 

\vspace{-1mm}
\subsection{Two-stage Efficient Generation}
\label{sec:method-twostage}

\begin{figure}[t]
\setlength{\abovecaptionskip}{0.mm}
    \centering
    \begin{minipage}{0.45\textwidth}
        \centering
        \includegraphics[width=\linewidth]{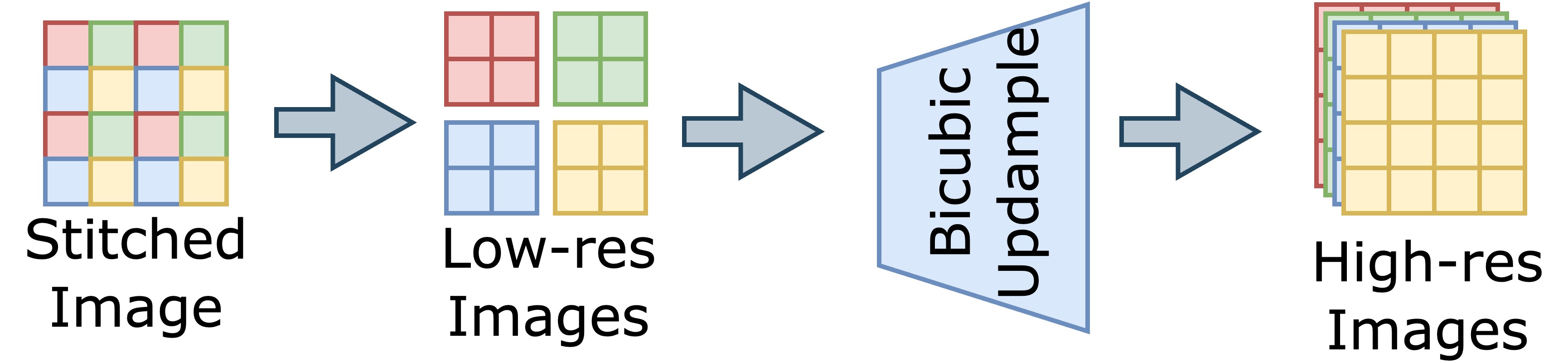} 
        \caption{The trade-off between inference latency and quality at different cache ratios is illustrated. The result shows that a cache ratio of 50\% offers the most balanced improvement.}
        \label{fig:evaluation-cache}
    \end{minipage}
    \hfill
    \begin{minipage}{0.45\textwidth}
        \centering
        \includegraphics[width=\linewidth]{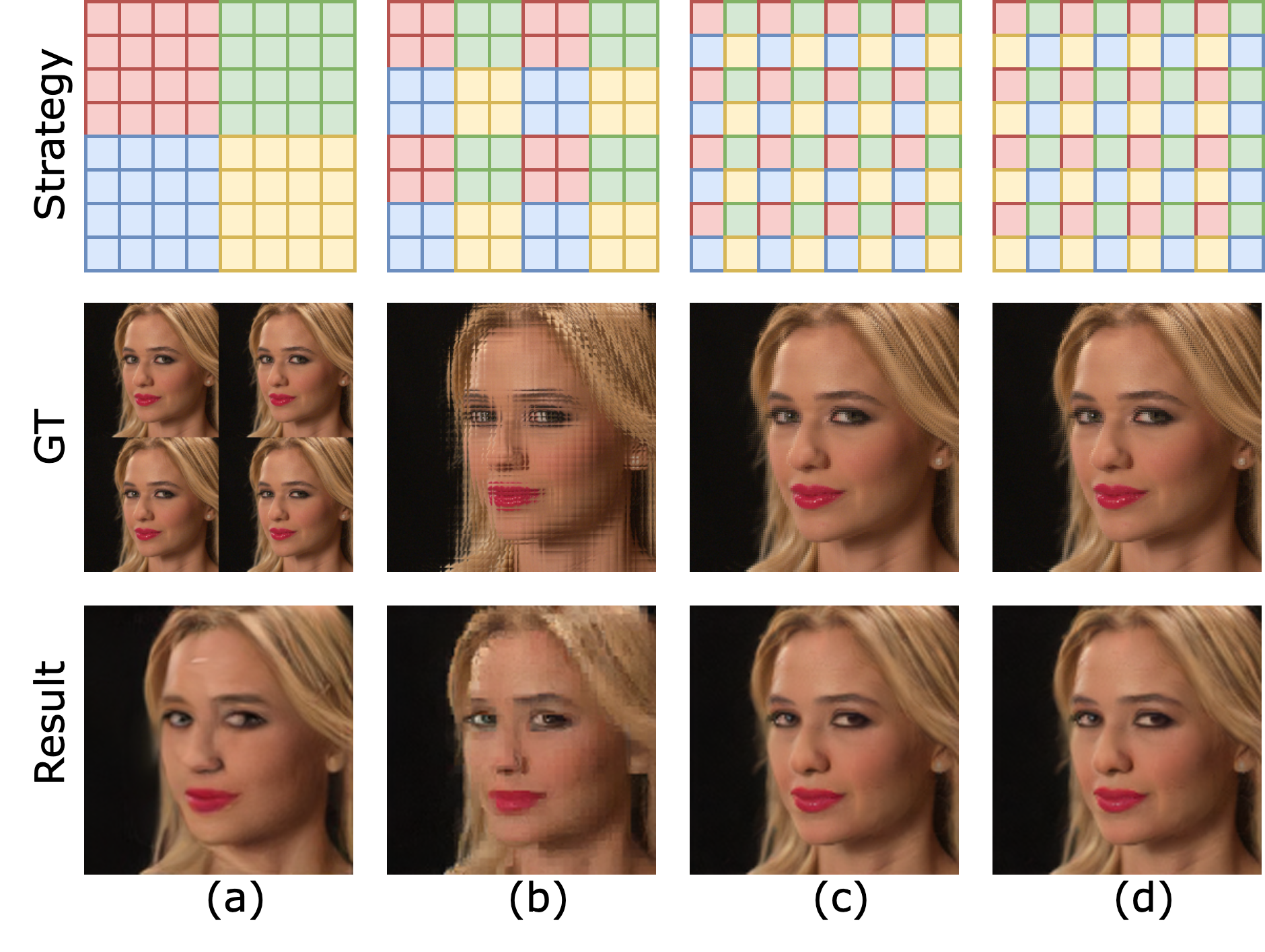}
        \caption{Comparison of different stitching strategies. (a) Direct stitching leads to uneven quality and flickering. (b) Patch-based stitching introduces high-frequency blocking artifacts. (c)(d) Pixel-by-pixel stitching improves spatial consistency, and our adaptive scheme selects the better arrangement from (c) or (d) to optimize fidelity and stability.}
        \label{fig:method-stich}
    \end{minipage}
    \hfill
    \begin{minipage}{0.45\textwidth}
        \centering
        \includegraphics[width=\linewidth]{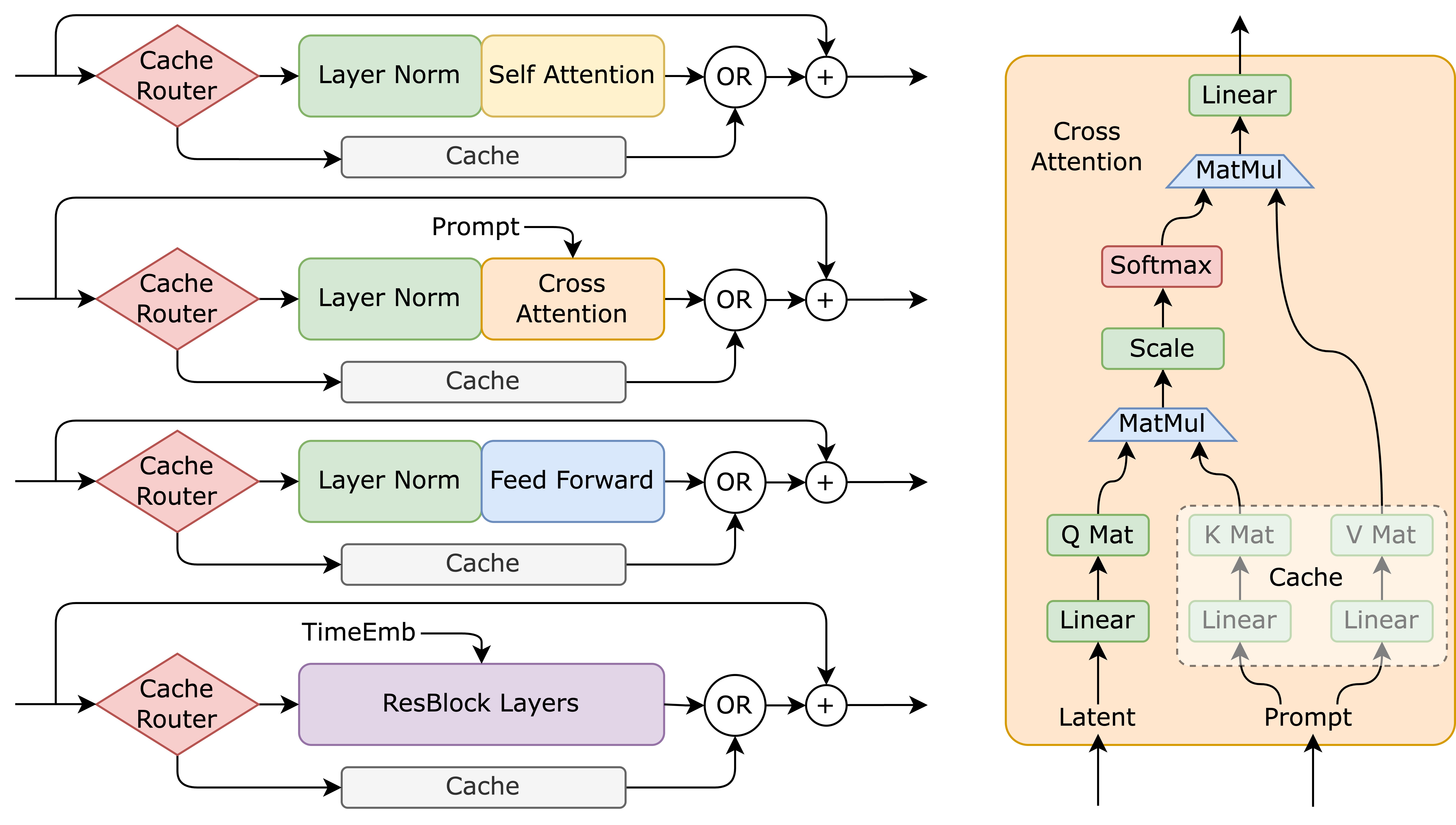}
        \caption{Main modules and cache strategy in the denoising network. Left: four residual block types (self-attention, cross-attention, feed-forward, ResBlock) allow caching of $\Delta$x due to inter-frame similarity. Right: The linear layers in the cross-attention module enable lossless caching by reusing K and V matrices. Thanks to prompt linear interpolation and the linear nature of certain operations, we replace computationally expensive linear layers with efficient caching and linear scaling.}
        \label{fig:method-cache-pipeline}
    \end{minipage}
    \vspace{-5mm}
\end{figure}

To reduce image generation computational cost, we developed a two-stage generation approach, as shown in Fig. \ref{fig:method-stich-pipeline}. First, the network generates a low-resolution image, which is then upsampled by a lightweight upsampling module. Experiments show that even the simplest cubic-bilinear upsampling can yield satisfactory images, as seen in Fig. \ref{fig:method-stich}(c). Additionally, as linear interpolation, which aligns image resolution with the phone's display resolution, is part of the image display process on mobile devices in scaleToFill content mode, the time consumed by this step can be disregarded. As shown in Fig. \ref{fig:motivation-resolution-time}, the iPhone 16 Pro Max can generate a 128-resolution video at about 27 FPS by using an unmodified SD 2.1 turbo. 

To address the problem of Stable Diffusion's inability to generate high-quality low-resolution images, we employed an image-stitching approach. The simplest form is direct stitching, as depicted in Fig. \ref{fig:method-stich}(a). Yet, this stitching method yields images that are rarely seen by the stable diffusion model, and the detailed features are confined locally. As a result, the final fitting effect is comparable to that of directly using a 128-resolution image. Besides, as stable diffusion exhibits variable performance at different image regions, the quality of the four frames fluctuates if this form of stitching is used. This fluctuation gives rise to flickering in the final video, considerably degrading the subjective viewing quality. 

To better harness the capabilities of the stable diffusion model and enhance stability, we explored pixel-by-pixel stitching, as depicted in Fig. \ref{fig:method-stich}(c). Given the smooth motion between video frames, this stitching approach can generate images just like an independent 256 $\times$ 256-resolution image, substantially enhancing the frame fitting quality. However, substantial object motion leads to high-frequency aliasing artifacts, posing challenges for image fitting. Consequently, we expanded the pixel arrangement scheme to include multiple configurations, as depicted in Fig. \ref{fig:method-stich}(d), which modifies the pixel arrangement. The optimal arrangement, having the lowest relative frequency and most closely resembling a single frame, is selected by employing the VGG Loss\cite{johnson2016perceptual} between the stitched frame and the component frames. Following the selection process, a two-bit data identifier is used to specify the stitching scheme, resulting in no substantial increase in the data volume for transmission. We also attempted to stitch images in the form of small patches, as shown in Fig. \ref{fig:method-stich}(b). However, the stitched images exhibited an even higher frequency, and the images reconstructed using this ground truth have very obvious blocking artifacts. 

Another significant benefit of image stitching is its potential to further cut down bandwidth requirements. According to Promptus, during intra-frame compression, the rank of $U$ and $V$ is $8$, and for inter-frame compression, a group of 10 frames can effectively convey motion and detail information in a video. Our frame-stitching approach enables the simultaneous generation of four frames. As a result, while maintaining the same transmission data volume, we can adopt a higher rank and a shorter key prompt interval. Tests reveal that a 5-frame group with a rank of 16 yields the highest-quality images generated by the prompt. 

\vspace{-3mm}

\subsection{Fine-grained Inter-frame Caching}
\label{sec:method-cache}


After conducting a detailed study of the stable diffusion network structure, we discovered that when utilizing the pipeline of Promptus, certain operations are redundant. Specifically, these redundant operations can be eliminated by leveraging cache, thereby enhancing the inference speed without any compromise in quality. Additionally, for non-linear operations, we observed a high degree of similarity among the intermediate variables within the network. This similarity can be exploited to reduce the computational load. Therefore, in the following sections, we will elaborate on the implementation of lossless cache and lossy cache strategies. 

\noindent\textbf{Lossless KV cache}: The structure of the cross-attention is as shown in the right side of Fig. \ref{fig:method-cache-pipeline}. Because our intermediate prompts are derived by linear interpolation, we can cache the K and V matrices. In the subsequent frame, we perform only linear scaling, instead of a complex Linear neural network. 

\noindent\textbf{Lossy cache}: Even though the computations in other segments of the network are nonlinear, we observe that, owing to linear interpolation of prompt, the output of some layers changes in a relatively smooth manner and can be approximately characterized by a linear transformation. As shown in the left side of Fig. \ref{fig:method-cache-pipeline}, the network is mainly composed of self-attention, cross-attention, feed-forward, and res blocks. All these components of the network can be abstracted as $y = x + \Delta x$. Consequently, we are able to cache $\Delta x$. Additionally, the determination of our cache strategy occurs during the generation of prompts. So these strategy configurations are transmitted to the mobile device via a few additional bits of data. The mobile device operates solely based on the received caching strategy. As a result, the optimal cache method can be selected in advance, and there is no need for the mobile device to store unnecessary historical variables, thus saving on memory overhead.
\vspace{-5mm}

\subsection{System-level Optimizations}
\noindent\textbf{Collaborative Training}: The newly introduced two-stage upsampling generation and lossy inter-frame cache both lead to quality loss. Given that prompts are obtained through iterative optimization using the gradient-descent method, we adopt a collaborative training approach to reduce the impact of such quality deterioration. Our co-training is divided into two phases. In the first phase, we train the prompt by using the complete non-cached network to generate 256-resolution images with the Tiny Decoder and upsampling strategy. Then in the second phase, we use the topK method to find the most similar $\Delta x$s in the U-Net and reuse them as inter-frame cache. We then finetune the prompt using the cached network and with other elements remaining the same. Furthermore, we observed that when applying the caching strategy without prompt fine-tuning, increasing the cache ratio does not degrade image quality but instead reduces inter-frame motion. At 100\% caching, the video becomes nearly static, leading to a great quality descent. To address this, rather than simply caching $\Delta x$, we introduce trainable parameters $k$ and $b$. These parameters are first initialized randomly. During the second-stage collaborative training, $k$ and $b$ are jointly fine-tuned with the prompt to compensate for the motion reduction caused by high caching ratios.

\noindent\textbf{Residual Flow}: When the quality of generated videos still remains unsatisfactory quality after joint training, we have the option to transmit an additional Residual Flow. However, the transmission of the residual flow will consume extra bandwidth. Therefore, we can make a slight reduction in the intra-frame compression rank. A 1-unit reduction in rank saves 13,212 bps of bandwidth, which we can use to transmit an extra residual flow stream to offset image quality degradation. Given that we have already transmitted the key frame stream, the residual stream only needs to encode motion vectors and residuals, obviating the need for large I frames. By modifying the H.265 codec, we achieved an LPIPS improvement of 0.012 with a rank reduction of 8.

\noindent\textbf{ANE Hardware Optimization}: Apple's Neural Engine (ANE) is a specialized hardware component designed to accelerate neural network computations on Mobile devices. However, it has strict requirements regarding the dimensions and formats of the data it processes. To fully harness the power of the ANE for our Stable Diffusion network, we replace the linear layers with 1D convolutions, leading to a significant reduction of the inference time. 

\noindent\textbf{A light-weight tiny decoder} Moreover, we replaced the time-consuming decoder module with the Tiny AutoEncoder for Stable Diffusion(TAESD). TAESD is a tiny, distilled version of Stable Diffusion's VAE. It uses the same "latent API" and can decode Stable Diffusion's latent into full-size images at nearly zero cost. After this replacement, the time consumption was reduced by 10 times. Since we trained the prompt in an end-to-end manner, the quality of the finally fitted images did not decline significantly.

\section{EVALUATIONS}
\label{sec:evaluation}

\subsection{Experimental Setting}

\noindent\textbf{Hardware}: We conducted our experiments on an iPhone 16 Pro Max, leveraging its Neural Engine with a computational capacity of nearly 45 Tera Operations Per Second (TOPS). 

\noindent\textbf{Dataset}: We use the UVG dataset to evaluate the quality of video fitting. The UVG dataset is a benchmark in video processing research. It encompasses diverse video content including natural and man-made scenes. All videos were preprocessed to a standardized format, with each frame cropped to a resolution of 512 $\times$ 512 and a frame rate of 30 FPS. 

\noindent\textbf{Baselines}: In the domain of low-bitrate video transmission, it is mainly categorized into traditional codecs, neural-enhanced streaming, and semantic communication. Among them, \textbf{traditional codecs} such as H.264~\citep{264}, H.265~\citep{265}, VP8~\citep{bankoski2011technical}, and VP9~\citep{mukherjee2015technical} primarily reduce the bitrate by spatial and temporal redundancy reduction, including motion estimation, transform coding, and entropy coding. However, despite their sophisticated designs, traditional codecs face significant limitations in extremely low-bandwidth scenarios.\textbf{Token-based methods} like VQGAN~\citep{li2023reparo,esser2021taming} quantize the video into tokens for sending. Specifically, the sender first uses a VAE~\cite{kingma2014auto} to map the video to latent variables and then quantifies these latent variables into tokens using a learned codebook. The receiver can reconstruct the tokens into a video using the same codebook and VAE. NAS~\citep{yeo2018neural} is the most representative \textbf{neural-enhanced streaming}, which uses DNNs for video super-resolution. Unlike other super-resolution-based methods~\citep{park2023omnilive,zhou2022cadm}, one of its major contributions is training a content-aware DNN for each video and sending it along with the video. Since the DNN is overfitted to the video, it can avoid quality loss caused by the domain gap. 

\noindent\textbf{Visual Metrics}: We adopt a Learned Perceptual Image Patch Similarity (LPIPS)~\citep{zhang2018unreasonable} to evaluate perceptual quality since it has been proven to more effectively reflect human subjective perceptions of video quality compared to traditional metrics like PSNR\cite{rabbani2002image} and SSIM\cite{wang2004image}. 



\begin{figure}[t]
\setlength{\abovecaptionskip}{0.mm}
    \centering
    \begin{minipage}{0.25\textwidth}
        \centering
        \includegraphics[width=\linewidth]{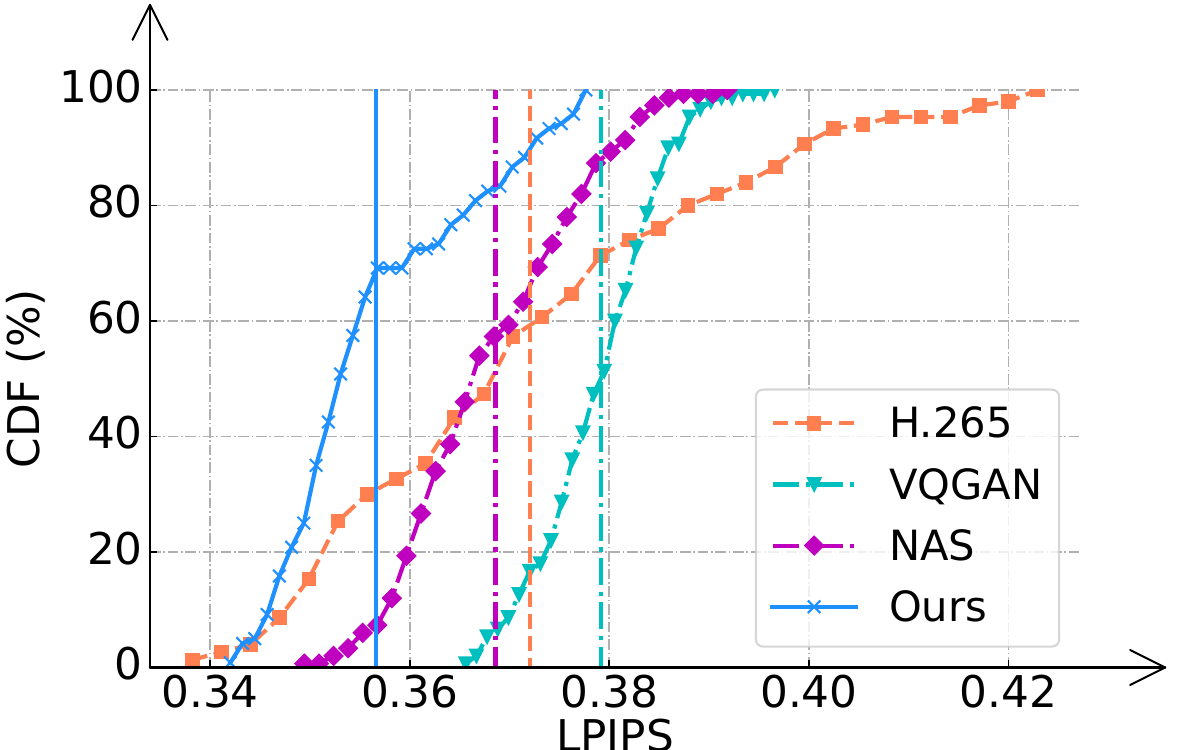} 
        \caption{Our work outperforms existing video codec methods in video quality, achieving an average LPIPS improvement of 0.016 compared to H.265, and reducing the number of severely distorted frames by 60\% when compared to VQGAN.}
        \label{fig:evaluation-quality}
    \end{minipage}
    \hfill
    \begin{minipage}{0.2\textwidth}
        \centering
        \includegraphics[width=\linewidth]{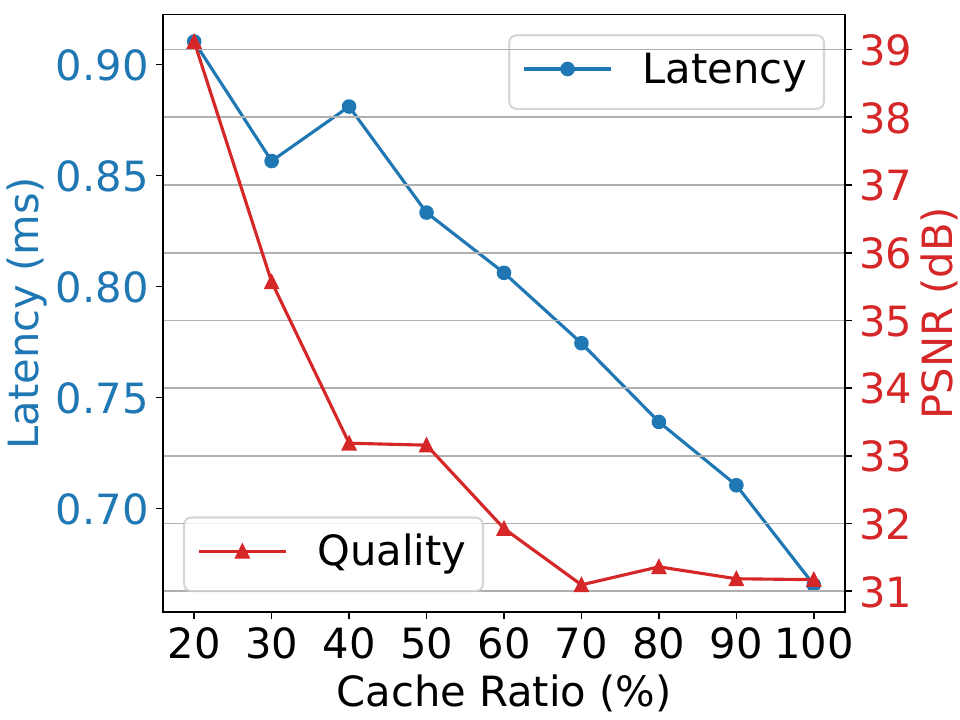}
        \caption{The trade-off between inference latency and quality at different cache ratios is illustrated. The result shows that a cache ratio of 50\% offers the most balanced improvement.}
        \label{fig:evaluation-cache}
    \end{minipage}
    \vspace{-4mm}
\end{figure}

\subsection{Cache Ratio Trade-off}

To investigate the trade-off between quality and computational efficiency, we conducted an experiment evaluating different cache ratios in our model. We systematically tested cache ratios ranging from 0\% to 100\% in 10\% increments, measuring both the generation quality and the corresponding time delay for each configuration, shown in Fig. \ref{fig:evaluation-cache}.

As the cache ratio gradually increases to 100\%, the running time continuously decreases, while the magnitude of the quality decline gradually flattens. The result shows that a cache ratio of 50\% can significantly shorten the running time without a substantial decline in image quality, achieving the most balanced improvement in quality and time. 

\vspace{-2mm}
\subsection{Video Quality}

This section demonstrates the video quality. We uniformly set the bandwidth to 280 kbps. However, to enable VQGAN and NAS to run in real-time on mobile phones, we appropriately reduced the resolution of the images generated by the two. Then, we used the same bicubic linear interpolation as in our work to upsample the images to the resolution of the original video before conducting the quality evaluation.

First, our method achieves better video quality. For example, in terms of the average LPIPS value, our approach reaches 0.356, which demonstrates a clear advantage. Second, our method can significantly reduce the ratio of severely distorted frames. For example, our method has 40\% of the frames with LPIPS values higher than 0.354. In comparison, H.265 only has 70\% of its frames fall below the same LPIPS threshold. As for NAS, only 96\% of its frames meet this criteria. Notably, 100\% frames generated by VQGAN have a quality inferior to 0.354. These improvements are attributed to Promptus's compression capabilities, along with our methods for maintaining quality during acceleration.
\vspace{-1mm}

\subsection{Ablation on Quality and Lantency}

In this ablation experiment, we systematically investigated the impact of speed and image quality of various components on the performance of the Stable Diffusion 2.1 turbo optimized for ANE. The components under study included Residual Flow, Collaborative Training, lossy cache, two-stage generation strategy, and TinyDecoder. Results are listed in Tab. \ref{tab:cache_ratio}

\begin{table}[ht]
\setlength{\abovecaptionskip}{0.mm}
    \centering
    \caption{Ablation study results of main modules in this work.}
    \small
    \begin{tabular}{c|c|c|c|c|c|c}
        \toprule
        Residual & Train & Cache & TwoStage & Tiny & LPIPS & FPS \\
        \midrule
        \xmark  & \xmark & \xmark & \xmark & \xmark & 0.229 & 3.13 \\
        \xmark  & \xmark & \xmark & \xmark & \cmark & 0.268 & 4.79 \\
        \xmark  & \xmark & \xmark & \cmark & \cmark & 0.362 & 38.81 \\
        \xmark  & \xmark & \cmark & \cmark & \cmark & 0.371 & 46.58 \\
        \xmark  & \cmark & \cmark & \cmark & \cmark & 0.368 & 46.58 \\
        \cmark  & \cmark & \cmark & \cmark & \cmark & 0.356 & 42.61 \\
        \bottomrule
    \end{tabular}
    \label{tab:cache_ratio}
    \vspace{-3mm}
\end{table}

Initially, the baseline model, the ANE-optimized Stable Diffusion 2.1 turbo, achieved a frame rate of 3.1 FPS with an LPIPS metric of 0.229. When the TinyDecoder was introduced, the generation speed increased to 4.79 FPS. However, the LPIPS value rises to 0.268. Subsequently, the incorporation of the Two-stage generation strategy led to a remarkable increase in the frame rate, soaring to 38.81 FPS. Nevertheless, the image quality deteriorated further, with the LPIPS value reaching 0.362. By integrating the existing modules during the prompt fitting process and conducting end-to-end Collaborative Training, we managed to slightly improve the image quality. The LPIPS value comes to 0.368. Finally, after reducing the rank of the prompt and adding the Residual Flow, we achieved an average LPIPS of 0.356. Although the generation speed slightly decreased to 44.6 FPS due to the additional operations of video decoding and image overlay, the overall performance in terms of both speed and quality reached a relatively balanced state.  

In conclusion, each component in the ablation experiment has a distinct impact on the performance of the model. The two-stage generation strategy is effective in improving generation speed but at the expense of image quality. Collaborative Training can enhance image quality, and the introduction of Residual Flow can achieve a more balanced performance in terms of both speed and quality.
\vspace{-2mm}

\section{LIMITATIONS}
\label{sec:limitations}

Our method exhibits two primary limitations. 

First, when adjacent frames differ drastically such as in scene cuts, extremely fast motion, or flickering, the image stitching strategy may produce high-frequency artifacts in the resulting image. These artifacts pose a challenge for stable diffusion to accurately fit the visual content. To mitigate this, we fall back to fitting only a single 256$\times$256 reference key frame, while relying on residual propagation to recover the remaining frames. 

Second, our approach requires optimizing the prompt embeddings via gradient descent, which involves thousands of iterations. As a result, the optimization for the first frame typically takes around 30 minutes to converge. Subsequent frames benefit from initialization using the previous frame's prompt, reducing convergence time to approximately 3 minutes per frame. However, since our method targets video-on-demand scenarios where generation can be performed offline, the relatively long training time is generally acceptable.

\vspace{-1mm}

\section{CONCLUSION}
\label{sec:conclusion}


This paper presents PromptMobile, an efficient acceleration framework tailored for on-device Promptus. Specifically, we propose (1) a two-stage efficient generation framework to reduce computational cost by 8.1x, (2) a fine-grained inter-frame caching strategy to reduce redundant computations by 16.6\%, and (3) system-level optimizations to further enhance efficiency. The evaluations demonstrate that compared with the original Promptus, PromptMobile achieves a 13.6x increase in image generation speed. Compared with other streaming methods, PromptMobile achieves an average LPIPS improvement of 0.016, reducing 60\% of severely distorted frames. 
\vspace{-1mm}

\section{ACKNOWLEDGMENT}

We thank the anonymous reviewers for their valuable feedback. This work was sponsored by the NSFC grant(62431017), Bytedance Grant(CT20241126107484). We gratefully acknowledge the support of State Key Laboratory of Media Convergence Production Technology and Systems, Key Laboratory of Intelligent Press Media Technology. Zongming Guo is the corresponding author.


\bibliographystyle{ACM-Reference-Format}
\bibliography{sample-base}


\begin{thebibliography}{22}


\ifx \showCODEN    \undefined \def \showCODEN     #1{\unskip}     \fi
\ifx \showISBNx    \undefined \def \showISBNx     #1{\unskip}     \fi
\ifx \showISBNxiii \undefined \def \showISBNxiii  #1{\unskip}     \fi
\ifx \showISSN     \undefined \def \showISSN      #1{\unskip}     \fi
\ifx \showLCCN     \undefined \def \showLCCN      #1{\unskip}     \fi
\ifx \shownote     \undefined \def \shownote      #1{#1}          \fi
\ifx \showarticletitle \undefined \def \showarticletitle #1{#1}   \fi
\ifx \showURL      \undefined \def \showURL       {\relax}        \fi
\providecommand\bibfield[2]{#2}
\providecommand\bibinfo[2]{#2}
\providecommand\natexlab[1]{#1}
\providecommand\showeprint[2][]{arXiv:#2}

\bibitem[264(2024)]%
        {264}
 \bibinfo{year}{2024}\natexlab{}.
\newblock \bibinfo{title}{H.264}.
\newblock
\newblock
\shownote{\url{https://www.itu.int/rec/T-REC-H.264}}.


\bibitem[265(2024)]%
        {265}
 \bibinfo{year}{2024}\natexlab{}.
\newblock \bibinfo{title}{H.265}.
\newblock
\newblock
\shownote{\url{https://www.itu.int/rec/T-REC-H.265}}.


\bibitem[sd(2024)]%
        {sd}
 \bibinfo{year}{2024}\natexlab{}.
\newblock \bibinfo{title}{Stable Diffusion}.
\newblock
\newblock
\shownote{\url{https://stability.ai/}}.


\bibitem[Bankoski et~al\mbox{.}(2011)]%
        {bankoski2011technical}
\bibfield{author}{\bibinfo{person}{Jim Bankoski}, \bibinfo{person}{Paul Wilkins}, {and} \bibinfo{person}{Yaowu Xu}.} \bibinfo{year}{2011}\natexlab{}.
\newblock \showarticletitle{Technical overview of VP8, an open source video codec for the web}. In \bibinfo{booktitle}{\emph{2011 IEEE International Conference on Multimedia and Expo}}. IEEE, \bibinfo{pages}{1--6}.
\newblock


\bibitem[Chen et~al\mbox{.}(2023)]%
        {chen2023speed}
\bibfield{author}{\bibinfo{person}{Yu-Hui Chen}, \bibinfo{person}{Raman Sarokin}, \bibinfo{person}{Juhyun Lee}, \bibinfo{person}{Jiuqiang Tang}, \bibinfo{person}{Chuo-Ling Chang}, \bibinfo{person}{Andrei Kulik}, {and} \bibinfo{person}{Matthias Grundmann}.} \bibinfo{year}{2023}\natexlab{}.
\newblock \showarticletitle{Speed is all you need: On-device acceleration of large diffusion models via gpu-aware optimizations}. In \bibinfo{booktitle}{\emph{Proceedings of the IEEE/CVF Conference on Computer Vision and Pattern Recognition}}. \bibinfo{pages}{4651--4655}.
\newblock


\bibitem[Esser et~al\mbox{.}(2021)]%
        {esser2021taming}
\bibfield{author}{\bibinfo{person}{Patrick Esser}, \bibinfo{person}{Robin Rombach}, {and} \bibinfo{person}{Bjorn Ommer}.} \bibinfo{year}{2021}\natexlab{}.
\newblock \showarticletitle{Taming transformers for high-resolution image synthesis}. In \bibinfo{booktitle}{\emph{Proceedings of the IEEE/CVF conference on computer vision and pattern recognition}}. \bibinfo{pages}{12873--12883}.
\newblock


\bibitem[Face(2023)]%
        {huggingface_minisd}
\bibfield{author}{\bibinfo{person}{Hugging Face}.} \bibinfo{year}{2023}\natexlab{}.
\newblock \bibinfo{title}{miniSD-diffusers}.
\newblock \bibinfo{howpublished}{\url{https://huggingface.co/lambdalabs/miniSD-diffusers}}.
\newblock


\bibitem[Johnson et~al\mbox{.}(2016)]%
        {johnson2016perceptual}
\bibfield{author}{\bibinfo{person}{Justin Johnson}, \bibinfo{person}{Alexandre Alahi}, {and} \bibinfo{person}{Li Fei-Fei}.} \bibinfo{year}{2016}\natexlab{}.
\newblock \showarticletitle{Perceptual losses for real-time style transfer and super-resolution}. In \bibinfo{booktitle}{\emph{Computer Vision--ECCV 2016: 14th European Conference, Amsterdam, The Netherlands, October 11-14, 2016, Proceedings, Part II 14}}. Springer, \bibinfo{pages}{694--711}.
\newblock


\bibitem[Kingma and Welling(2014)]%
        {kingma2014auto}
\bibfield{author}{\bibinfo{person}{Diederik~P Kingma} {and} \bibinfo{person}{Max Welling}.} \bibinfo{year}{2014}\natexlab{}.
\newblock \showarticletitle{Auto-Encoding Variational Bayes}.
\newblock \bibinfo{journal}{\emph{Journal of Machine Learning Research (JMLR)}} \bibinfo{volume}{15}, \bibinfo{number}{1} (\bibinfo{year}{2014}), \bibinfo{pages}{1929--1958}.
\newblock


\bibitem[Li et~al\mbox{.}(2023)]%
        {li2023reparo}
\bibfield{author}{\bibinfo{person}{Tianhong Li}, \bibinfo{person}{Vibhaalakshmi Sivaraman}, \bibinfo{person}{Lijie Fan}, \bibinfo{person}{Mohammad Alizadeh}, {and} \bibinfo{person}{Dina Katabi}.} \bibinfo{year}{2023}\natexlab{}.
\newblock \showarticletitle{Reparo: Loss-resilient generative codec for video conferencing}.
\newblock \bibinfo{journal}{\emph{arXiv preprint arXiv:2305.14135}} (\bibinfo{year}{2023}).
\newblock


\bibitem[Mukherjee et~al\mbox{.}(2015)]%
        {mukherjee2015technical}
\bibfield{author}{\bibinfo{person}{Debargha Mukherjee}, \bibinfo{person}{Jingning Han}, \bibinfo{person}{Jim Bankoski}, \bibinfo{person}{Ronald Bultje}, \bibinfo{person}{Adrian Grange}, \bibinfo{person}{John Koleszar}, \bibinfo{person}{Paul Wilkins}, {and} \bibinfo{person}{Yaowu Xu}.} \bibinfo{year}{2015}\natexlab{}.
\newblock \showarticletitle{A technical overview of vp9—the latest open-source video codec}.
\newblock \bibinfo{journal}{\emph{SMPTE Motion Imaging Journal}} \bibinfo{volume}{124}, \bibinfo{number}{1} (\bibinfo{year}{2015}), \bibinfo{pages}{44--54}.
\newblock


\bibitem[Park et~al\mbox{.}(2023)]%
        {park2023omnilive}
\bibfield{author}{\bibinfo{person}{Seonghoon Park}, \bibinfo{person}{Yeonwoo Cho}, \bibinfo{person}{Hyungchol Jun}, \bibinfo{person}{Jeho Lee}, {and} \bibinfo{person}{Hojung Cha}.} \bibinfo{year}{2023}\natexlab{}.
\newblock \showarticletitle{Omnilive: Super-resolution enhanced 360 video live streaming for mobile devices}. In \bibinfo{booktitle}{\emph{Proceedings of the 21st Annual International Conference on Mobile Systems, Applications and Services}}. \bibinfo{pages}{261--274}.
\newblock


\bibitem[Rabbani and Wang(2002)]%
        {rabbani2002image}
\bibfield{author}{\bibinfo{person}{M. Rabbani} {and} \bibinfo{person}{P. Wang}.} \bibinfo{year}{2002}\natexlab{}.
\newblock \bibinfo{booktitle}{\emph{Image Compression: Fundamentals, Techniques, and Applications}}.
\newblock \bibinfo{publisher}{SPIE Press}.
\newblock


\bibitem[Sandvine(2024)]%
        {global-phenomena24}
\bibfield{author}{\bibinfo{person}{Sandvine}.} \bibinfo{year}{2024}\natexlab{}.
\newblock \bibinfo{title}{2024 Global Internet Phenomena Report}.
\newblock
\newblock
\shownote{\url{https://www.sandvine.com/global-internet-phenomena-report-2024}}.


\bibitem[{Statista}(2023)]%
        {statista_mobile_video}
\bibfield{author}{\bibinfo{person}{{Statista}}.} \bibinfo{year}{2023}\natexlab{}.
\newblock \bibinfo{title}{Mobile video in the United States - statistics \& facts}.
\newblock
\urldef\tempurl%
\url{https://www.statista.com/topics/2725/mobile-video-in-the-united-states}
\showURL{%
\tempurl}
\newblock
\shownote{Accessed: 2025-03-14}.


\bibitem[Wang et~al\mbox{.}(2004)]%
        {wang2004image}
\bibfield{author}{\bibinfo{person}{Z. Wang}, \bibinfo{person}{A.~C. Bovik}, \bibinfo{person}{H.~R. Sheikh}, {and} \bibinfo{person}{E.~P. Simoncelli}.} \bibinfo{year}{2004}\natexlab{}.
\newblock \showarticletitle{Image quality assessment: From error visibility to structural similarity}.
\newblock \bibinfo{journal}{\emph{IEEE Transactions on Image Processing}} \bibinfo{volume}{13}, \bibinfo{number}{4} (\bibinfo{year}{2004}), \bibinfo{pages}{600--612}.
\newblock


\bibitem[Wimbauer et~al\mbox{.}(2024)]%
        {wimbauer2024cache}
\bibfield{author}{\bibinfo{person}{Felix Wimbauer}, \bibinfo{person}{Bichen Wu}, \bibinfo{person}{Edgar Schoenfeld}, \bibinfo{person}{Xiaoliang Dai}, \bibinfo{person}{Ji Hou}, \bibinfo{person}{Zijian He}, \bibinfo{person}{Artsiom Sanakoyeu}, \bibinfo{person}{Peizhao Zhang}, \bibinfo{person}{Sam Tsai}, \bibinfo{person}{Jonas Kohler}, {et~al\mbox{.}}} \bibinfo{year}{2024}\natexlab{}.
\newblock \showarticletitle{Cache me if you can: Accelerating diffusion models through block caching}. In \bibinfo{booktitle}{\emph{Proceedings of the IEEE/CVF Conference on Computer Vision and Pattern Recognition}}. \bibinfo{pages}{6211--6220}.
\newblock


\bibitem[Wu et~al\mbox{.}(2024)]%
        {wu2024promptus}
\bibfield{author}{\bibinfo{person}{Jiangkai Wu}, \bibinfo{person}{Liming Liu}, \bibinfo{person}{Yunpeng Tan}, \bibinfo{person}{Junlin Hao}, {and} \bibinfo{person}{Xinggong Zhang}.} \bibinfo{year}{2024}\natexlab{}.
\newblock \showarticletitle{Promptus: Can Prompts Streaming Replace Video Streaming with Stable Diffusion}.
\newblock \bibinfo{journal}{\emph{arXiv preprint arXiv:2405.20032}} (\bibinfo{year}{2024}).
\newblock


\bibitem[Xu et~al\mbox{.}(2024)]%
        {xu2024accelerating}
\bibfield{author}{\bibinfo{person}{Xiaoxia Xu}, \bibinfo{person}{Yuanwei Liu}, \bibinfo{person}{Xidong Mu}, \bibinfo{person}{Hong Xing}, {and} \bibinfo{person}{Arumugam Nallanathan}.} \bibinfo{year}{2024}\natexlab{}.
\newblock \showarticletitle{Accelerating mobile edge generation (MEG) by constrained learning}.
\newblock \bibinfo{journal}{\emph{arXiv preprint arXiv:2407.07245}} (\bibinfo{year}{2024}).
\newblock


\bibitem[Yeo et~al\mbox{.}(2018)]%
        {yeo2018neural}
\bibfield{author}{\bibinfo{person}{Hyunho Yeo}, \bibinfo{person}{Youngmok Jung}, \bibinfo{person}{Jaehong Kim}, \bibinfo{person}{Jinwoo Shin}, {and} \bibinfo{person}{Dongsu Han}.} \bibinfo{year}{2018}\natexlab{}.
\newblock \showarticletitle{Neural adaptive content-aware internet video delivery}. In \bibinfo{booktitle}{\emph{13th USENIX Symposium on Operating Systems Design and Implementation (OSDI 18)}}. \bibinfo{pages}{645--661}.
\newblock


\bibitem[Zhang et~al\mbox{.}(2018)]%
        {zhang2018unreasonable}
\bibfield{author}{\bibinfo{person}{Richard Zhang}, \bibinfo{person}{Phillip Isola}, \bibinfo{person}{Alexei~A Efros}, \bibinfo{person}{Eli Shechtman}, {and} \bibinfo{person}{Oliver Wang}.} \bibinfo{year}{2018}\natexlab{}.
\newblock \showarticletitle{The unreasonable effectiveness of deep features as a perceptual metric}. In \bibinfo{booktitle}{\emph{Proceedings of the IEEE conference on computer vision and pattern recognition}}. \bibinfo{pages}{586--595}.
\newblock


\bibitem[Zhou et~al\mbox{.}(2022)]%
        {zhou2022cadm}
\bibfield{author}{\bibinfo{person}{Qihua Zhou}, \bibinfo{person}{Ruibin Li}, \bibinfo{person}{Song Guo}, \bibinfo{person}{Peiran Dong}, \bibinfo{person}{Yi Liu}, \bibinfo{person}{Jingcai Guo}, {and} \bibinfo{person}{Zhenda Xu}.} \bibinfo{year}{2022}\natexlab{}.
\newblock \showarticletitle{Cadm: Codec-aware diffusion modeling for neural-enhanced video streaming}.
\newblock \bibinfo{journal}{\emph{arXiv preprint arXiv:2211.08428}} (\bibinfo{year}{2022}).
\newblock


\end{thebibliography}










\end{document}